\documentclass[aps,prd,twocolumn,superscriptaddress,floats,amsfonts,amsmath,amssymb]{revtex4}
\usepackage{graphicx,epsf}
\usepackage[active]{srcltx}
\usepackage{bm}
\usepackage[]{latexsym}

\newcommand{\bea}{\begin{eqnarray}}\newcommand{\eea}{\end{eqnarray}}
\newcommand{\be}{\begin{eqnarray}}\newcommand{\ee}{\end{eqnarray}}
\newcommand{\ba}{\begin{array}}\newcommand{\ea}{\end{array}}
\newcommand{\bit}{\begin{itemize}}\newcommand{\eit}{\end{itemize}}
\newcommand{\ben}{\begin{enumerate}}\newcommand{\een}{\end{enumerate}}

\usepackage{float}

\usepackage{hyperref}
\hypersetup{pdfauthor={huchiayu},pdftitle={},
            pdfsubject={},pdfcreator={pdfLaTeX with hyperref (\today)}}

\begin{document}
\restylefloat{figure}

\title{A GPU-based Calculation Method for Near Field Effects of Cherenkov Radiation Induced by Ultra High Energy Cosmic Neutrinos}

\author{Chia-Yu Hu}
\email{r96244004@ntu.edu.tw}
\affiliation{Graduate Institute of Astrophysics, National Taiwan University, Taipei, Taiwan 10617}
\affiliation{Leung Center for Cosmology and Particle Astrophysics (LeCosPA), National Taiwan University, Taipei, Taiwan, 10617}
\author{Chih-Ching Chen}
\email{chen.chihching@gmail.com}
\affiliation{Graduate Institute of Astrophysics, National Taiwan University, Taipei, Taiwan 10617}
\affiliation{Leung Center for Cosmology and Particle Astrophysics (LeCosPA), National Taiwan University, Taipei, Taiwan, 10617}
\author{Pisin Chen}
\email{pisinchen@phys.ntu.edu.tw}
\affiliation{Graduate Institute of Astrophysics, National Taiwan University, Taipei, Taiwan 10617}
\affiliation{Leung Center for Cosmology and Particle Astrophysics (LeCosPA), National Taiwan University, Taipei, Taiwan, 10617}
\affiliation{Department of Physics, National Taiwan University, Taipei, Taiwan 10617}
\affiliation{Kavli Institute for Particle Astrophysics and Cosmology, SLAC National Accelerator Laboratory, Menlo Park, CA 94025, U.S.A.}

\begin{abstract}
The radio approach for detecting the ultra-high energy cosmic neutrinos has become a mature field.
The Cherenkov signals in radio detection are originated from the charge excess of particle showers due to Askaryan effect.
The conventional way of calculating the Cherenkov pulses by making Fraunhofer approximation fails 
when the sizes of the elongated showers become comparable with the detection distances.
We present a calculation method of Cherenkov pulses based on the finite-difference time-domain (FDTD) method, 
and attain a satisfying effeciency via the GPU-acceleration.
Our method provides a straightforward way of the near field calculation, 
which would be important for ultra high energy particle showers, 
especailly the electromagnetic showers induced by the high energy leptons
produced in the neutrino charge current interactions.

\vspace{3mm}

\noindent {\footnotesize PACS numbers:}
\end{abstract}

\maketitle

\section{Introduction}
Cosmic neutrinos, as a probe of the universe to the highest energy regime, are wonderful in many respects.
Due to their extremely small interaction cross section, 
they can penetrate through galactic infrared (IR) and cosmic microwave background (CMB) photons, 
while photons of energy above 10 TeV would be attenuated.
Furthermore, being uncharged, they propagate along straight lines and therefore are able to point directly back to their sources,
while protons or other charged particles would be deflected by the magnetic field in the universe.

Ultra-high energy cosmic rays (UHECRs) have been observed up to $\approx$ $10^{19.6}$ eV.
The source of such amazingly energetic events have remained a mystery.
Above this energy scale, UHECRs interact with CMB photons through the Greisen-Zatsepin-Kuzmin(GZK) processes~\cite{gzk_process}.
The GZK cut-off of the cosmic ray energy spectrum has been first observed 
by the High Resolution Fly's Eye Experiment~\cite{cutoffHiRes}
and later confirmed by the Pierre Auger Observatory~\cite{cutoff}, 
so the corresponding GZK neutrinos are almost guaranteed to exist.
Nevertheless, none of these have been observed so far. 
Detecting the GZK neutrinos provides critical informations 
for unraveling the mystery of the origin and evolution of the cosmic accelerators, 
and will be one of the utmost tasks in the coming decade~\cite{pisin_whitepaper}.

One promising way of detecting UHE neutrinos is the radio approach.
When an ultra-high energy cosmic neutrino interacts with ordinary matters on the Earth, 
it would lead to a hadronic debris, either by charged current or neutral current.
The former also produces a lepton with corresponding flavor.
Both the high energy leptons and the hadronic debris induce particle showers.
As proposed by Askaryan in the 1960's~\cite{Askaryan}, 
the high energy particle shower develops in a dense medium would have net negative charges.
This charge imbalance appears as a result of the knocked-off electrons being part of the shower,
as well as the positrons in the shower annihilating with the electrons of the medium.
The net charges of the showers, typically $20 \%$ of total shower particles, 
serve as a source emitting the Cherenkov radiations when they travel in the medium.
The sizes of the showers are quite localized (tens of cm in radial and few meters in longitudinal development) 
compared to those develope in the air (km scale),
and therefore result in coherent radiations for wavelengths longer than the shower sizes.
The corresponding coherent wavelength turns out to be in the radio band, from hundreds of MHz to few GHz.
This Askaryan effect has been confirmed in a series of experiments at Stanford Linear Accelerator Center (SLAC), 
using different dense media such as silica sand, rock salt and ice~\cite{aska1, aska2, aska3, aska4}.


\section{Coherent Cherenkov Pulses}
In the radio detection experiment, 
the signals come from the Cherenkov radiations of the net charges in the shower.
The key concept which makes this technique possible is the coherent emission.
In fact, the Cherenkov radiation is a broad band emission and the intensity increases as frequency.
For a single charged particle,
the Cherenkov signal in the radio band should be the weakest in the spectrum.
It is the compact size of the shower that makes radio signal so special. 
The coherent emission greatly enhences the signal strength in the radio band.

The electric field of Cherenkov radiations can be calculated by solving the inhomogeneous Maxwell equations, 
as it has been demonstrated in the paper of Zas, Halzen, and Stanev~\cite{ZHS}.
The vector potential can be obtained by the Green's function method:
\small{
\begin{equation}
	\vec{A}_{\omega}(\vec{x}) 
	= \frac{1}{4 \pi \epsilon_0 {\rm c}^2}
	\int\!\!d^{3}\!x'\,\frac{\exp{(ik|\vec{x}-\vec{x}\,'|)}}
	{|\vec{x}-\vec{x}\,'|}\int\!\!dt'\,\exp{(i\omega t')}
	\vec{J}(t',\vec{x}\,') \label{eq:GreenMethod}
\end{equation}
}
\normalsize
where $k$ is the wavenumber,
$\vec{x}$ the position of the detector,
$\vec{x}\,'$ the position of the shower particles,
and $\vec{J}$ the current sources.
Adopting the Fraunhoffer approximation
\small{
\begin{equation}
	\frac{ \exp{(ik|\vec{x}-\vec{x}\,'|)} }{|\vec{x}-\vec{x}\,'|}
	\approx   \frac{ \exp{(ik|\vec{x}|-i k\hat{x}\cdot\vec{x}\,')} }{R} ~,
	\label{eq:fraun_appr}
\end{equation}
}
\normalsize
where $R$ is the absolute value of $\vec{x}$,
the integration in Eq.(~\ref{eq:GreenMethod}) can be greatly simplified and 
therefore enhances the computational effeciency in the Monte Carlo simulation~\cite{ZHS, AZ_1d, AZ_unified, AZ_thinned}.
The validity of this approximation relies on several length scales:
the detection distance ($R$), the spatial size of the shower ($l$) and the wavelengths of interest ($\lambda$).
The Fraunhoffer approximation works well under the condition
\small{
\begin{equation}
	\frac{l^2}{\lambda R}~\rm{sin}^2\theta \ll 1
	\label{eq:farfield_condition}
\end{equation}
}
\normalsize
where $\theta$ is the observational angle between the shower axis and the observational direction.

However, for the ultra-high energy showers the longitudinal development is longer,
especially for the electromagnetic showers that suffer from the Landau-Pomeranchuk-Migdal (LPM) suppression~\cite{LPM, lpm_klein}.
Electromagnetic showers can be produced by the charge current generated leptons.
For a electromagnetic shower of EeV-scale energy,
The impact of LPM effect on the shower development has been investigated by Monte Carlo simulations~\cite{Niess, Bolmont, AZ_thinned}.
Electromagnetic showers of primary energies $10^{20}$ eV can be extended to about 200-m long with great fluctuations.
In such cases, the far field condition cannot be satisfied for distance up to several kilometers,
while the typical detection distance for ground array detectors is about 1 km due to the attenuation length of radio signals in ice.
Under these circumstances, the Fraunhoffer approximation is clearly invalid,
and one has to deal with the complicated integration in Eq.(~\ref{eq:GreenMethod}).

In the paper of Buniy and Ralston~\cite{BR},
the correction has been made by the saddle point approximation,
while it still cannot cope with the extreme cases for $R \sim l$, the near field regime.
We handle this problem by a numerical method based on first principle
so that the near field radiations can effeciently obtained.
Although far field radiations would be more time consuming, 
it is not our focus in this paper.

Hadronic showers, on the other hand, are less affected by LPM effect~\cite{alvarez98, acorne},
since the sources of electromagnetic components in hadronic showers are the decay of neutral pions,
which tend to be interact with matters instead of decay at energy above 6.7 PeV.
The far field condition is well fulfilled for hadronic showers in most parctical cases.


\section{Numerical Method}
Numerical algorithms for calculating electromagnetic fields have been existed for decades. 
However, it was not until the recent rapid growth of computational power that this approach became wildly adopted.
Among all the existing algorithms, 
the finite-difference time-domain (FDTD hereafter) method has several advantages:

\begin{itemize}
\item
It is exceptionally simple to be implemented by computer programs.
\item 
It is a time-domain approach that is well suitable for an impulse signal.
A broad band impulse can be calculated in one single run.
\item 
The algorithm itself is inherently parallel and the effeciency can be largely improved via parellel computing.
\end{itemize}

The idea of FDTD was first proposed by Yee in the 1960's~\cite{Yee},
and has been in use for many years for the electromagnetic impulse modeling.
Like most of the numerical finite difference methods, 
the space are discretized into small grids and fields are calculated on each grid by solving Maxwell equations.
Adopting a special lattice arrangment (known as the Yee lattice), 
the E-field and H-field are staggered in both space and time and can be calculated in a leapfrog time-marching way.

\subsection{Algorithm}

The Maxwell curl equations in differential forms are
\small{
\begin{equation}
	\nabla \times \mathbf{E} = -\mu      \frac{\partial \mathbf{H}}{\partial t} ~,
\end{equation}
\begin{equation}
	\nabla \times \mathbf{H} =  \epsilon \frac{\partial \mathbf{E}}{\partial t} + \sigma \mathbf{E} ~.
\end{equation}
}
\normalsize

\begin{figure}[htbp]
	\begin{center}
	\includegraphics[width=8cm]{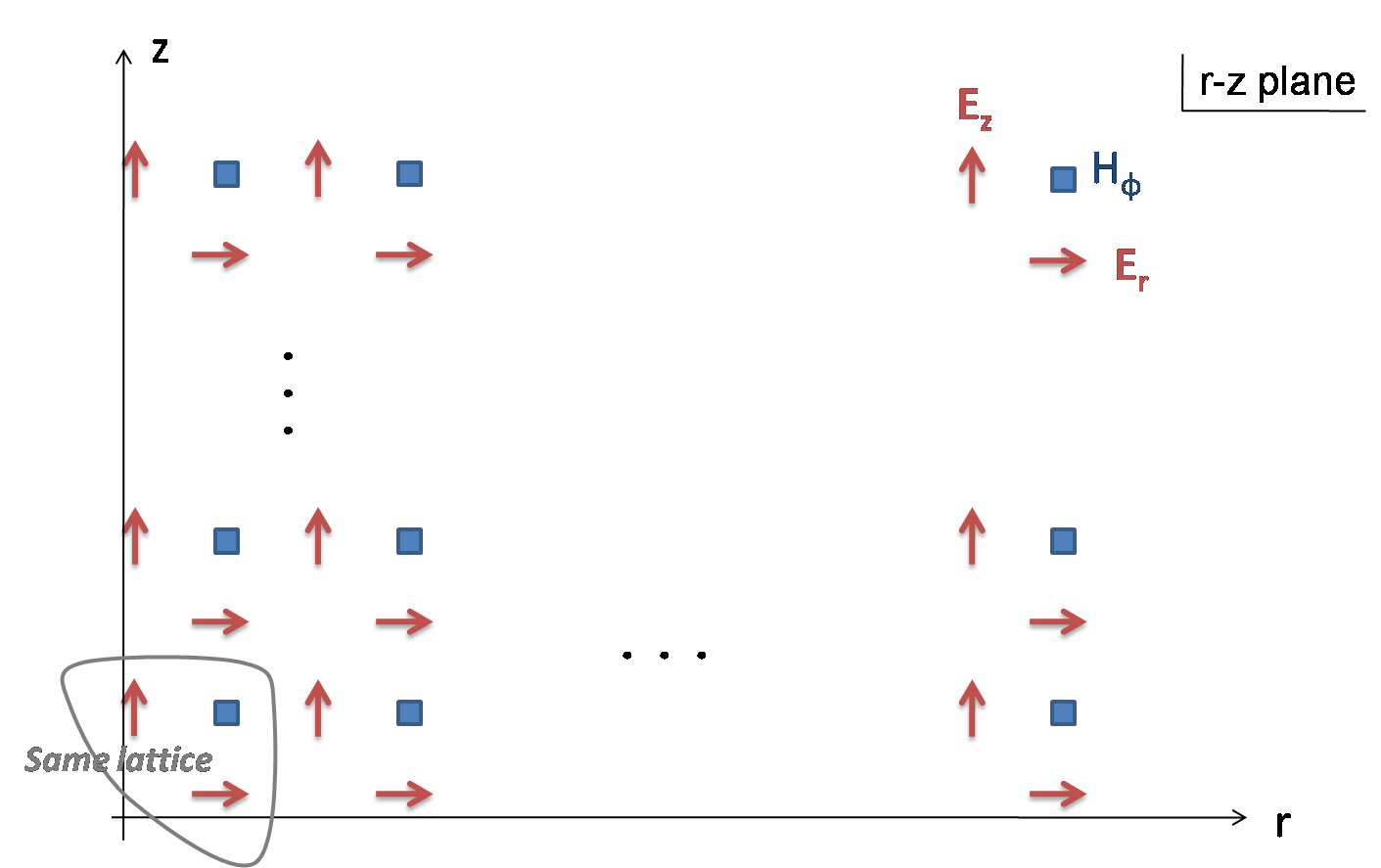}
	\caption{The lattice configuration we adopted.}
	\label{fig:cylind_grid}
	\end{center}
\end{figure}

The FDTD method approximates derivatives by finite differences. 
The central difference is adopted to achieve 2nd order accuracy in both spatial and temporal derivatives.
We assume cylindrical symmetry along the shower axis (defined as z-axis), 
and therefore all the derivatives with respect to $\phi$ vanish.
In addition, due to the polarization property of Cherenkov radiations, 
the $H_r, H_z, E_\phi$ components also vanish.
This can save large amounts of computer memories as well as calculation time.
Figure(~\ref{fig:cylind_grid}) shows the configuration of the lattice under these assumptions.
Maxwell equtions in a cylindrical coordinate are reduced as:
\small{
\begin{eqnarray}
	-\frac{\partial H_\phi}{\partial z} = \epsilon \frac{\partial E_r}{\partial t} + \sigma E_r \label{eq:fdtd_Er} \\
	\frac{1}{r} \frac{\partial (rH_\phi)}{\partial r} = \epsilon \frac{\partial E_z}{\partial t} + \sigma E_z \label{eq:fdtd_Ez} \\
	\frac{\partial E_r}{\partial z} - \frac{\partial E_z}{\partial r} = -\mu \frac{\partial H_\phi}{\partial t} \label{eq:fdtd_Hphi}
\end{eqnarray}
}
\normalsize
We can use Eq.(~\eqref{eq:fdtd_Er}) and Eq.(~\eqref{eq:fdtd_Ez}) to update the E-field, 
and then use Eq.(~\eqref{eq:fdtd_Hphi}) to update the H-field.
The spatial and temporal grid sizes have been chosen such that the numerical stability is satisfied
and the numerical dispersion is controlled at an acceptable level~\cite{Courant, Taflove}.

\subsection{Simulation Setup}
In order not to lose the focus on the RF calculation, 
we simply assume a shower model for the longitudinal development rather than invoking the Monte Carlo packages.
For electromagnetic showers, 
the well known Nishimura-Kamata-Greisen (NKG) parametrization formula~\cite{NKG} describing the number of shower particles is
\small{
\begin{equation}
	N_e(X) = \frac{0.31}{ \sqrt{\ln{E_0/E_c}} } \exp \left[\left(1-\frac{3}{2}\ln{s}\right)\frac{X}{X_R}\right]
	\label{eq:nkg_formula}
\end{equation}
}
\normalsize
where $E_0$ is the energy of the primary particle and $E_c$ the critical energy,
$X$ the slant depth, $X_{max}$ depth as shower reaching its maximum number,
$s$ the (dimensionless) shower age defined as $s = 3X/(X+X_{max})$.
The NKG formula can be fit by a Gaussian distribution as:
\small{
\begin{equation}
	N_e(z) = N_{{\rm max}} \exp(-\frac{z^2}{2l^2})
	\label{eq:showerModel}
\end{equation}
}
\normalsize
where $N_{{\rm max}}$ is the particle number at shower maximum
and $l$ is the longitudinal shower length.
The NKG formula for $E_0 = 10$ TeV has $l \sim 2$ m in ice.
For charge distribution in a snapshot,
we assume the Gaussian distribution in both radial ($r$) and longitudinal ($z$) directions:
\small{
\begin{equation}
	n(z,r) = \frac{N_e}{(2\pi)^{1.5}\sigma_z\sigma_r^2} 
	\exp \left(-\frac{(z-X)^2}{2\sigma_z^2}\right) 
	\exp \left(-\frac{r^2}{2\sigma_r^2}\right)
\end{equation}
}
\normalsize
where $z$ and $r$ are the cylindrical coordinates in the unit of g/cm$^{2}$,
and $\sigma_z$ and $\sigma_r$ are the standard deviation of the distribution in $z$ and $r$ direction respectively.
We choose $\sigma_r$ = 5 cm in our simulations according to the Moliere radius in ice.
The $\sigma_z$ is generally even smaller than the Moliere radius 
and thus has no significant effect on the radiation pattern.

\begin{figure}[htbp]
	\begin{center}
	\includegraphics[width=6.5cm]{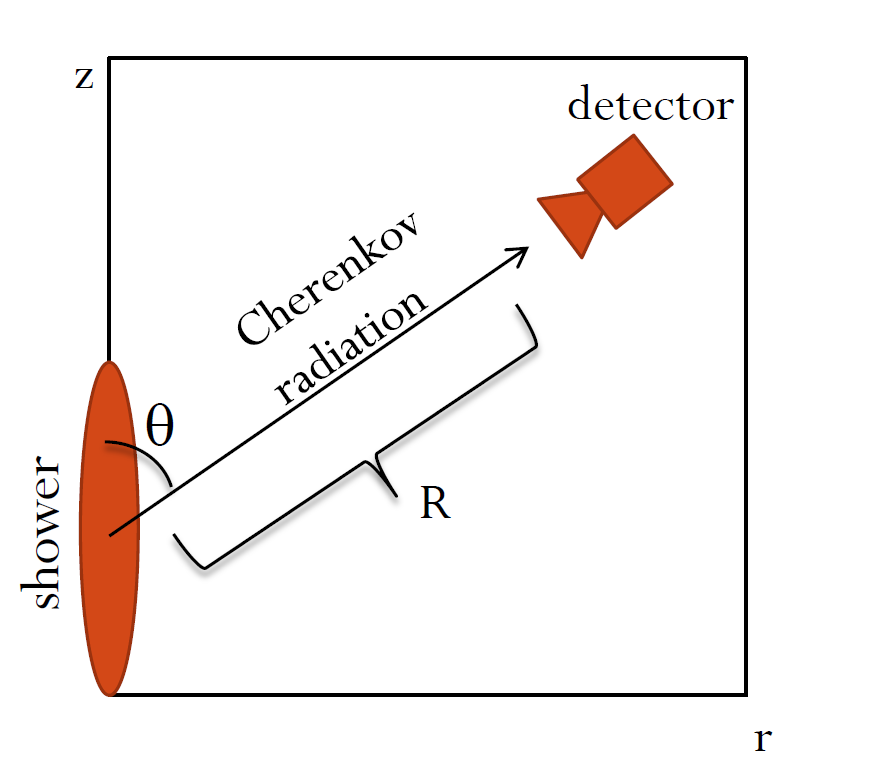}
	\caption{Cartoon for our simulation set up. }
	\label{fig:setup}
	\end{center}
\end{figure}

Figire~(\ref{fig:setup}) is a cartoon that demonstrates the setup.
The shower travels in the +z direction emitting Cherenkov radiations.
We define a spherical coordinate whose origin lies on the intersection of the z axis and the shower maximum.
Choosing the desired detector positions by varying detection angle ($\theta$) and detection distance ($R$), 
and record the electric field values as time evolves.
After one single run, we can have all the simulation results we want.
Here one can see the merit of using a time-domain calculation method.

\section{Performance Improvement via the GPU Parallelization}

Recently, graphical processing units (GPU) have became more and more important in the field of high performance computation. 
Under the needs in the computer game markets, the GPUs are currently under booming developments.
The GPUs have become programmable via the Compute Unified Device Architecture (CUDA) provided by NVIDIA, 
and have been implemented in various scientific areas such as molecular dynamics, gravitational N-body simulations and lattice QCD
~\cite{MD, sapporo, cudanbody, gamer, lattQCD}.
CUDA is an extension of C language, one of the most popular high level languages in the world.
Programmer familiar with C language can utilize GPU computations by simply calling the functions from CUDA.
The general parallelization strategies can be found in the CUDA Programming Guide (available on their webpage).

Because of the multicore architecture, 
GPUs are ideal for implementing parallel algorithms.
The FDTD method is therefore the ideal candidate to benefit from GPU computing.
The texture memories provide the possibility to realize the memory cache speedup.
Since the texture memories in CUDA are read-only,
we bind the 1D cuda array to E-field and H-field alternatively, 
and write back to global memories for the unbound E/H-field.
The basic steps are like the following:

\begin{enumerate}
\item
Allocate global memories to store E-field and H-field.
\item 
Bind texture to H-field.
\item 
Calculate and store the updated E-field by reading the cached H-field.
\item
Unbind H-field.
\item
Bind texture to E-field.
\item 
Calculate and store the updated H-field by reading the cached E-field.
\item
Unbind E-field.
\item
Repeat steps (2) - (7).
\end{enumerate}

We have investigated the performances of our codes with one single NVIDIA GTX285 graphic card, 
which has 240 cores and 933 floating point operations per second (GFlOPS) of theoretical peak performance.
The following table shows the performances of our codes with different total grid numbers (N), 
compared to Intel Quad Core i7 920 at 2.66GHz (sequential code without CPU parallelization).
The GPU system is provided by 
the Center for Quantum Science and Engineering of National Taiwan University (CQSE).

\begin{table}[h]
\begin{center}

\begin{tabular}{ l | c c c c } \hline \hline

$N_{\rm{grid}}$		&$t_{\rm{CPU}}$	&$t_{\rm{GPU}}$	&speed up		&GPU performance\\
			&(sec)  			&(sec)  			&(${t_{\rm{CPU}}}/{t_{\rm{GPU}}}$)  &(GFLOPS)\\
\hline
$256^2$		&5.15				&0.040				&128.75			&57.0 \\
$512^2$		&34.94				&0.186				&187.85			&98.1\\
$1024^2$	&294.41				&1.167				&252.28			&125.1\\
$2048^2$	&2122.80			&8.372				&253.56			&139.5\\
$4096^2$	&15751.17			&67.617				&232.95			&138.2\\
$8192^2$	&120617.48			&681.082			&177.11			&109.8\\
\hline \hline

\end{tabular}
\end{center}
\caption{Performance of CPU and GPU codes.}
\renewcommand{\arraystretch}{1.5}
\label{tab:benchmark}
\end{table}

The algorithm of FDTD method is inherently parallel since each grid can be updated independently,
so the advantage of GPU systems can be maximally utilized in this algorithm.
The performance test shows a tremendous speed up using GPU parallelization, 
which allows the calculation to be done in very reasonable time,
with a cost-efficient computer resource.

\section{Results}

\subsection{At the Cherenkov Angle}
\begin{figure}[H]
	\begin{center}
	\includegraphics[width=9cm]{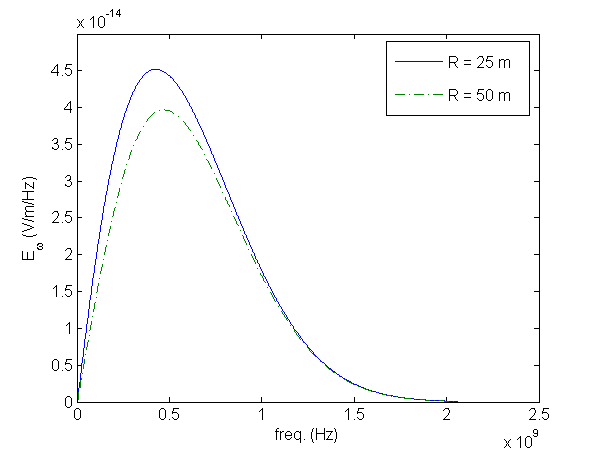}
	\caption{$E_\omega$ spectra at $R$ = 25 m and $R$ = 50 m with the latter scaled by a factor of $\sqrt{2}$; 
	these two spectra match well at high frequencies,
	implying a cylindrical behavior ($E_\omega$ $\propto$ $1/\sqrt{R}$).}
	\label{fig:near_match}
	\end{center}
\end{figure}

\begin{figure}[H]
	\begin{center}
	\includegraphics[width=9cm]{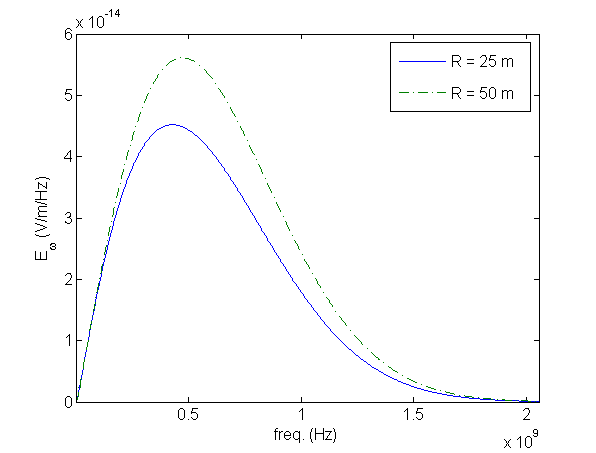}
	\caption{$E_\omega$ spectra at $R$ = 25 m and $R$ = 50 m with the latter scaled by a factor of 2; 
	these two spectra match well at low frequencies,
	implying a spherical behavior ($E_\omega$ $\propto$ $1/R$).}
	\label{fig:far_match}
	\end{center}
\end{figure}

First we compare the E-fields fixed at Cherenkov angle ($\theta_c$) with various distances ($R$).
The magnitude of the $E_\omega$ decreases as $R$ increases.
However, the decreasing speeds in different frequencies are not the same.
For example, Fig.(~\ref{fig:near_match}) shows the $E_\omega$ spectra at $R$ = 25 m and $R$ = 50 m, 
but with the latter multiplied by a factor of two.
These two spectra match well at high frequencies
while they deviate from each other at low frequencies.
In Fig.(~\ref{fig:far_match}),
the spectra at $R$ = 25 m and $R$ = 50 m are shown with the latter scaled by a factor of two.
In this case, the two spectra match well at low frequencies,
while they deviate at high frequencies.
According to these two examples,
one can see that the behavior in high frequencies suggests that $E_\omega$ $\propto$ $1/\sqrt{R}$, 
namely a cylindrical wave behavior.
On the other hand, 
the behavior in low frequencies suggests $E_\omega$ $\propto$ $1/R$,
a spherical wave behavior.

The different behaviors in different frequency regimes can be interpreted in a physical way.
It is a known fact that waves of higher frequencies are less likely to diffract.
Therefore, the high frequency waves in Cherenkov radiations are more confined in the $\theta$-direction, 
and their energies can only spread into the Cherenkov cone.
Their energies go like $1/R$ due to geometrical reason,
and hence $1/\sqrt{R}$ for the fields.
For the low frequency waves, 
diffraction allows another direction (the $\theta$-direction) for their energies to disperse, 
and therefore decrease faster.

The fact that the higher frequency regime decreases slower implies that 
there is a shift of the peak frequecy in different $R$.
The peak will migrate to the higher frequency regime as distance moves further away.
The shape of the spectrum is $R$-dependent, 
and a simple scaling relation of $R$ is no longer valid here.

\begin{figure}[H]
	\begin{center}
	\includegraphics[width=9cm]{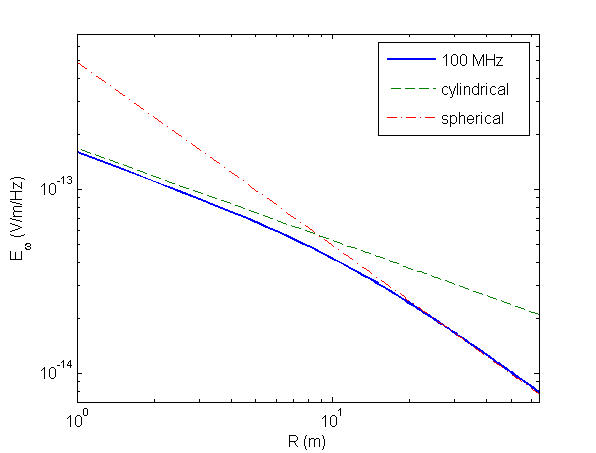}
	\caption{
	The $E_\omega$ - $R$ relation of 100MHz.
	The blue curve is the E-field.
	The green and red curve show the cylindrical ($E_\omega$ $\propto$ $1/\sqrt{R}$) 
	and spherical ($E_\omega$ $\propto$ $1/{R}$) behavoir respectively.
	The E-field goes like $1/\sqrt{R}$ in the near field regime,
	while approaches to ($1/R$)-behavoir in the far field regime.
	Note the smooth transition from near field to far field.
	}
	\label{fig:transition}
	\end{center}
\end{figure}

In principle, all waves will eventually diffract and behave like spherical waves no matter how high their frequencies are, if we set the distance $R$ infinitely large. Therefore, the terms "high" and "low" frequencies are only relative concepts. From the far field condition in Eq.(~\ref{eq:farfield_condition}) we can see 
that all three length scales couple with each other.
The waves start to diffract after they propagate to the distance large enough 
such that the far field condition is satisfied.
This character can be demonstrated more clearly if we plot the $E_\omega$ - $R$ relation with one single frequency.
Figure(~\ref{fig:transition}) shows how $E_\omega$ decreases with $R$.
At the distance very close to the shower, $E_\omega$ goes like $1/\sqrt{R}$.
As the distance increases, 
for large enough $R$,
the radiation can be viewed as a point source and thus have $E_\omega$ $\propto$ $1/{R}$.
We can see a smooth transition from cylindrical behavoir to spherical behavoir.
At $R$ large enough such that all the waves of different frequencies in the Cherenkov pulse reach the far field regime,
the shape of the spectrum is fixed and independent of $R$, and $R$ becomes just a normalization factor.

\subsection{Angular Distribution}
\begin{figure}[H]
	\begin{center}
	\includegraphics[width=9cm]{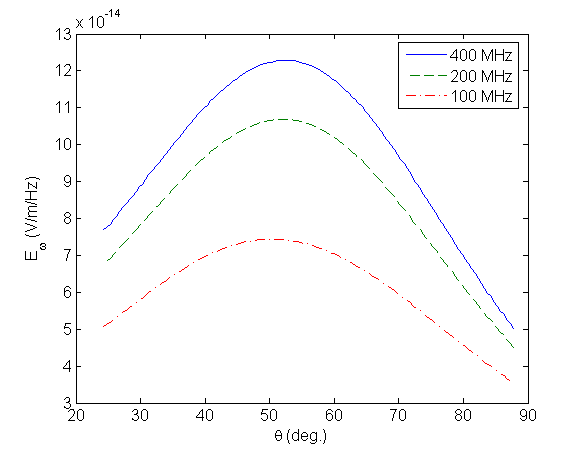}
	\caption{$E_\omega$ angular distributions for $R$ = 4 m (near field).
	The diffraction has not yet happened and the radiation pattern follows the shower image.
	}
	\label{fig:angular_4m}
	\end{center}
\end{figure}

\begin{figure}[H]
	\begin{center}
	\includegraphics[width=9cm]{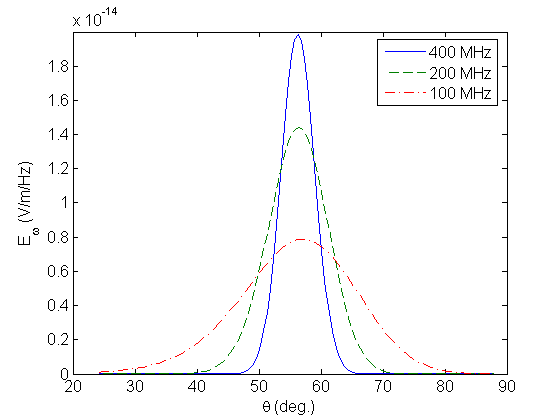}
	\caption{$E_\omega$ angular distributions for $R$ = 64 m (far field).
	The higher the frequency is, the narrower of its angular distribution is.
	Diffraction phenomenon is quite obvious here.
	}
	\label{fig:angular_64m}
	\end{center}
\end{figure}

\begin{figure}[H]
	\begin{center}
	\includegraphics[width=9cm]{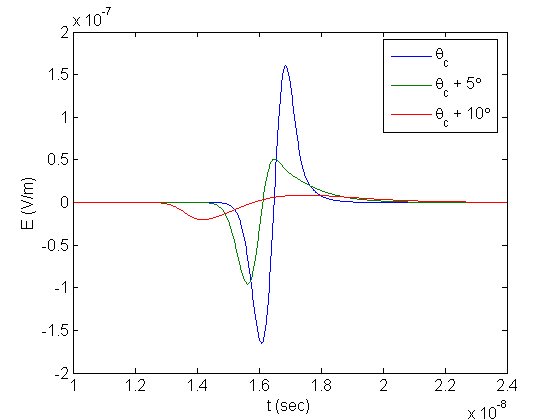}
	\caption{Time domain E-fields at different angles $\theta$ at $R$ = 32 m.
	The pulse is wilder at angles more away from $\theta_c$,
	implying the fact that lower frequency components diffract more.
	}
	\label{fig:timeDomain_32m}
	\end{center}
\end{figure}

The diffraction effect can also be seen in the angular distribution.
Figures(~\ref{fig:angular_4m}) and (~\ref{fig:angular_64m}) 
show the angular distributions of $E_\omega$ at $R$ = 4 m and $R$ = 64 m respectively.
For the case at $R$ = 4 m, the distance is too close for the waves to diffract. 
In fact, the radiation pattern in the near field (before diffraction happen) is just a fuzzy image of the radiating source.
We can see the waves of higher frequencies have stronger magnitudes, 
which is the inherent character of Cherenkov radiations.
Note the distribution is not symmetric on two sides, 
since the $\theta < \theta_c$ part is closer to the shower axis than the $\theta > \theta_c$ part is.
This asymmetry can be understood as we are using the spherical coordinate 
to describe a sysytem that is actually cylindrical symmetric.
For the case at $R$ = 64 m, the detection distance is long enough for waves diffract into the $\theta$-direction.
The lower the frequency is, the wilder of its angular distribution is, 
which is the standard property in diffraction.
In time domain, it can be seen in Figure(~\ref{fig:timeDomain_32m}) that the pulse at detection angle more away from $\theta_c$ has wilder width.
In principle, if the frequency goes to infinity, the angular distribution will be a delta function.
However the destructively interference of the lateral distribution suppress these high frequecy components.
The distribution now looks much more symmetric, 
since the differences of the distances to the shower axis between $\theta < \theta_c$ and $\theta > \theta_c$ parts are negligible compares to $R$.
Namely, as $R$ goes further, the system, originally cylindrical symmetric, becomes more and more spherical symmetric.

\subsection{Comparison}

\begin{figure}[H]
	\begin{center}
	\includegraphics[width=9cm]{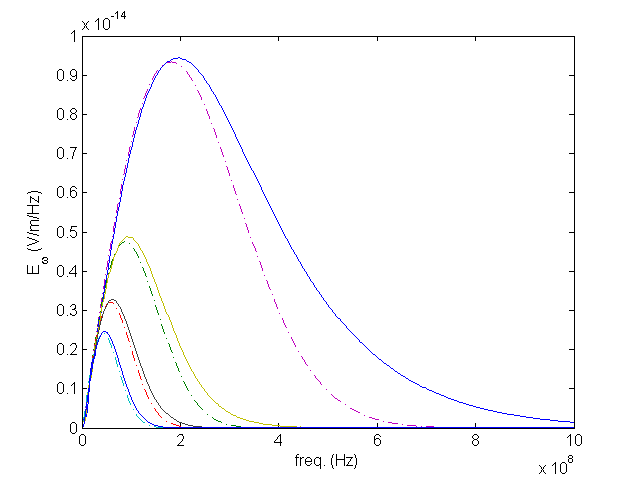}
	\caption{Comparison between the spectra of the far field formula and our simulation results at $R$ = 64 m.
	From top to bottom are the spaetra at 
	$\theta_c + 5^{\circ}$, $\theta_c + 10^{\circ}$, $\theta_c + 15^{\circ}$, $\theta_c + 20^{\circ}$ respectively.
	The solid curves are our results and the dashed curves are obtained from the far field formula.
	}
	\label{fig:matched}
	\end{center}
\end{figure}

We compare our results with the conventional far field formula.
The one dimensional approach in~\cite{AZ_1d} should be a reasonable approximation except at the Cherenkov angle.
Substituting our shower model in Eq.(~\ref{eq:showerModel}),
the E-field can be obtained:
\begin{align}
	\vec E(\omega,{\vec {\rm x}}) &=
	\frac{e}{4 \pi \epsilon_0 {\rm c}^2}~i \omega
	~\sin \theta~{{\rm e}^{ikR}\over R}~{\hat n_\perp}
	\int dz'~N_e(z')~{\rm e}^{i p z'}\nonumber \\
	&= 
	\frac{\sqrt{2\pi}e}{4 \pi \epsilon_0 {\rm c}^2}~i \omega
	~\sin \theta~{{\rm e}^{ikR}\over R}
	N_{{\rm max}} l
	~{\rm e}^{-\frac{l^{2} p^{2}}{2}}~{\hat n_\perp}
	\label{1d_appro}
\end{align}
where the parameter $p(\theta,\omega)= (1-n \cos \theta)~\omega /c$.
Figure (~\ref{fig:matched}) shows the comparison between the spectra of the far field formula and our simulation results 
at $R$ = 64 m and different $\theta$.
At lower frequency part they are in good agreements,
while at high frequency part there are significant differences between them.
It can be understood as the disagreement part has not yet reached the far field regime and thus decreases slower.
If we set larger $R$, the disagreement part will enter the far field regime and will match with the formula.

\section{Summary and Conclusions}
We have developed a numerical code to calculate the radiation patterns of Cherenkov signals 
from near field to far field based on the FDTD method.
By utilizing GPU parallel computation, 
the effeciency of the code can be greatly improved to a satisfying level on a comercial graphic card NVIDIA GTX285.
This will be useful in studying the signals originate from an elongated shower of its size comparable to the detection distance,
where the traditional Fraunhofer approximation does not apply.
Signals from the ultra-high energy electromagnetic showers induced by the electrons 
produced in neutrino charged-current interaction are the typical examples,
for they suffer from severe impact of the LPM effect.
Our result shows a smooth transition between near field and far field pattern.
In fact, the FDTD method is more suitable for the calculation of near field pattern
since the far field pattern may challange the computer resources.
The spectrum and the angular distribution of near field pattern have quite complicated $R$ dependences
instead of simple scalings in the case of far field.

In the cases of far field,
the angular distribution of signals induced by LPM-elongated showers are much narrower than the ordinary ones,
and the detection solid angle is considered to be small.
However the far field assumption neglects the shower size and treat it as a point source which is not fair.
A shower of hundred meters long would in fact generates signals spaneed also hundred meters, 
and is surely as possible to be detected as those with compact size.

The idea of using staggered grid configuration to solve the two coupled first-ordered PDEs is not limited to the electromagnetic problems.
Recently there are also applications of FDTD method in the acoustic simulations~\cite{ac_fdtd_1, ac_fdtd_2}
solving the coupled pressure fields and velocity fields.
It is possible to simulate signals in the neutrino detection experiments using acoustic approaches,
which is another potential field in UHE neutrino detection~\cite{acoustic, acousticBaikal}.

In the next generation neutrino detectors applying the ground array layout,
it is possible to simultaneously detect the hadronic shower and the electromagnetic shower 
that are induced by one single charged current neutrino interaction.
If both the hadronic shower and the electromagnetic shower can be correctly reconstructed,
it opens an opportunity to distinguish the electron neutrino from others~\cite{AZ_flavor},
since the two shower vertexes are nearly located at the same places.
However, the features of near field are very different from far field,
and will face some detection difficulties.
For example, the normal way to reconstruct the direction of incoming Cherenkov pulses by the arrival time differences between antennas
is based on the assumption that the shower is a point source, i.e. the far field assumption.
For a extended shower this assumption fails and therefore requires a new reconstruction method.
Any ground based neutrino detector has to take this near field effect into account
in order to reconstruct signals from extended showers.

\section{Acknowledgements}
We thank Melin Huang for useful discussions and Ting-Wai Chiu for the help in GPU-calculation.
This research is supported by Taiwan National Science Council(NSC) under Project No. NSC98-2811-M-002-501, No. NSC98-2119-M-002-001, 
the Center for Quantum Science and Engineering of National Taiwan University(NTU-CQSE) under Nos. 98R0066-65, 98R0066-69, 
and US Department of Energy under Contract No. DE-AC03-76SF00515.
We would also like to thank Leung Center for Cosmology and Particle Astrophysics for the support.

\end{document}